# Revisiting size effects in higher education research productivity[1]

**Abstract**

The potential occurrence of variable returns to size in research activity is a factor to be considered in choices about the size of research organizations and also in the planning of national research assessment exercises, so as to avoid favoring those organizations that would benefit from such occurrence. The aim of the current work is to improve on weaknesses in past inquiries concerning returns to size through application of a research productivity measurement methodology that is more accurate and robust. The method involves field-standardized measurements that are free of the typical distortions of aggregate measurement by discipline or organization. The analysis is conducted for 183 hard science fields in all 77 Italian universities (time period 2004-2008) and allows detection of potential differences by field.

# 1. Introduction

Policy-makers are ever more demanding of production efficiency in research activities and this in turn has required and stimulated much analysis of the research production function. With this, scholars have included examination of the possibility of economies of scale and of links between productivity and size of research groups.

To policy makers and managers, proof of returns to size would be of interest in decisions about sizing the production capacity of research organizations and units. The onset and diffusion of national evaluation exercises also means that the potential of variable returns to size should be considered in planning the actual assessment systems and the interpretation of results. If variable returns to size do occur then the organizations to be evaluated would first have to be reasonably subdivided by size and then compared within each size class. If this were not done then there would be a risk of rewarding or penalizing certain size groups over others on the basis of increasing or decreasing returns to size. Demonstration of constant returns to size would obviate any need for subdivision of the organizations by size[2].

Most studies on this issue have modeled research organizations as multi-business systems, active in fields of research, technology transfer and, for universities, also in teaching. Most studies concern economies of scale in research, attempting to quantify the reduction in average cost of product as level of output expands (Lewis and Dundar, 1999). Very few studies have specifically focused on the relationship between research productivity and size of the research unit. In either case, these studies have had contrasting results. Thus neither policy makers nor organization managers have clear indications to orient their decisions on this issue.

There are three principal difficulties in analyzing impact of staff size on research productivity. The first challenge concerns the necessity to separate research activities from teaching when analyzing university systems, which are the organizations most commonly of interest. The second concerns the separation of the size effects on productivity from those of other variables, such as quality of research staff, access to other production factors, localization etc. which vary from organization to organization. Concerning this problem, it is apparent that the larger the field of observation (in term of number of research organizations), the lesser is the probability of concentration of such positive or negative productivity factors in a single "size class" of organization. The third and most difficult challenge is how to identify and implement productivity measures that are accurate and robust, particularly in accounting for the problem that production functions are likely to be different in the various fields of research, as are the forms of codifying the outputs of research and the intensity of publication and citations.

With this work, the authors intend to revisit research about the relationship between staff size and productivity, overcoming past difficulties in examining the links. The field of observation, consisting of the 77 Italian universities active in the hard sciences over 2004-2008, is very large and has other characteristics that permit the separation of research production from education and the analysis of returns to size in research, free of size-related concentration of production factors other than labor. We apply a bibliometric methodology using indicators of productivity with standardization by subject category. The analysis, conducted for 183 different fields, permits consideration of varying production functions of the fields and detection of any relative differences in returns to size.

---

[2] As an example of such style, Italy's Triennial Research Evaluation exercise (2001-2003) subdivided research organizations into three groups based on size of their research staff (VTR, 2006).

The following section summarizes previous studies on returns to size and on economies of scale in higher education. Section 3 describes certain key characteristics of the Italian university system and the methodology we apply for the evaluation of research performance. Section 4 is divided in two parts: the first attempts to detect if "quality" of the labor production factor is uniformly distributed in organizations of different size (through the measurement of performance), or if it is instead significantly correlated with size of research unit; the second analyzes the relationship between productivity and size. The final section reviews the results and suggests some policy indications.

**2. Returns to size in higher education systems: literature review**

The literature on questions of efficiency in research and higher education systems tends to focus on economies of scale rather than on returns to size, although it would be reasonable to expect a certain alignment of results. However, studies concerning either question arrive themselves at contrasting results.

Studies in different parts of the world have attempted to detect the presence of economies of scale in research. Several studies in the USA suggest that there are indeed product-specific scale economies (Cohn et al., 1989; Dundar and Lewis, 1995; Koshal and Koshal, 1999; Laband and Lentz, 2003), but authors of other studies arrive at different conclusions (Adams and Griliches, 1998). The occurrence of economies of scale has also been shown for Europe (Izadi et al., 2001; Ducht-Brown, 2010) and in other parts of the world (Hashimoto and Cohn, 1998; Avrikan, 2001; Abbott and Doucouliagos, 2003; Longlong et al., 2009), but once again there are studies that arrive at contrasting conclusions (Worthington and Higgs, 2011).

Analysis of potential variable returns to size has received less attention in the literature. In general, such analysis focuses on the direct relationship between size and productivity. In their two works, Jordan et al. (1988, 1989) examine how size and form of organization affect research productivity, analyzing US academic departments first in economics only and then extending the study to 23 disciplines. Applying an OLS quadratic regression in which the dependent variable is the average number of publications per faculty member, they found support for the idea that "publishing activity increases with department size at a diminishing rate". To take account of differences across disciplines they applied OLS regression for each of the 23 separate disciplines, but they did not standardize the data by field, although they noticed there were substantial differences in publication rates across fields. Several years later, Golden and Carstensen (1992) applied a similar analysis to the same dataset and rejected the previous results, concluding that the impact of department size on productivity remains in doubt. Their conclusion was that the Jordan et al. model does not provide a sufficient data fit and that the coefficient for impact of size is statistically insignificant. Seglen and Aksnes (2000) analyzed 180 Norwegian microbiological research groups, defined on the basis of co-authorship, with at least one microbiological article during the period 1992-1996. Applying a correlation analysis, they concluded that the number of articles per capita was independent of group size. Bonaccorsi and Daraio (2005) analyzed France's INSERM in biomedical research and Italy's National Research Council in all scientific disciplines, showing returns to size that are generally constant or decreasing in all disciplines considered. Using a dataset derived from three

CNR reports and an INSERM database, and applying LOESS regression[3], the study correlates number of publications in international journals per capita and number of researchers in each institute, without accounting for differences in publication intensity across fields. The results indicated that in the majority of cases there is an absence of relation between these two variables.

All the works cited, while certainly useful in the advancement of knowledge on the subject, present a similar critical problem. Their examination lacks the necessary analysis at field level. This is essential, and brief consideration quickly explains why. Firstly, while the ultimate objective of research is always the advancement of science, the modalities of carrying out research are different from discipline to discipline and from field to field within the same discipline. A jurist draws on different production factors than does a biologist. Presumably the production function in juridical research is different than that in biology. Certainly the prevailing forms of codifying research are different: monographs and note to an appeal court judgment for the former, and articles in international refereed journals for the second. The overall universities, and other research organizations, are also never uniform in terms of their fields of scientific activity: some are specialized; others multidisciplinary, with varying extent of diversification. Thus, when conducting bibliometric evaluation, it is an unacceptable assumption to consider organizations as homogenous, as is done when carrying out the analyses at the aggregate level of entire organizations using an undifferentiated indicator of output for all. Even when analysis is restricted to the hard sciences there are important differences among disciplines in intensity of production of articles and in intensity of citation. This phenomenon is due only in part to the different numbers of journals indexed for the disciplines in the main bibliometric databases, such as Web of Science (WoS) or Scopus. In Table 1 we see, for each hard-science discipline, the notable differences in average research output (articles, article reviews, and conference proceedings) per capita, as indexed in WoS for Italian universities between 2004 to 2008.

It also unacceptable to conduct aggregate analyses at the level of scientific discipline. To provide an example concerning the sole discipline of biology, the field of Biochemistry shows an average research output per capita (2.16) that is more than double that for Plant sciences (0.97). And again regarding biology, an article falling in the field of Biochemistry receives an average of about 24 citations over eight years, while an article in Mycology receives about 7. Without standardizing by field it would be very difficult for a research group of mycologists to ever show productivity like that of biochemists. All bibliometricians agree on this point: measures based on aggregation at the discipline level, without field standardization, lead to unacceptable distortions in assessing productivity of research organizations (Moed, 2005; Abramo et al., 2008).

In light of these problems, the intention of the authors is to provide a much more convincing and broad answer to research questions about the link between the productivity of a research unit and its size, through an examination of the entire Italian university system and the use of an innovative measurement method for research productivity. The methodological details are described in the following section.

| Discipline | Total output | Research staff | Average output per year |
|---|---|---|---|
| Mathematics and computer sciences | 14,043 | 3,288 | 0.854 |
| Physics | 22,321 | 2,576 | 1.733 |

---

[3] Local regression (i.e. non-linear) method.

| | | | |
|---|---:|---:|---:|
| Chemistry | 24,629 | 3,241 | 1.520 |
| Earth sciences | 4,639 | 1,275 | 0.728 |
| Biology | 28,192 | 5,198 | 1.085 |
| Medicine | 50,603 | 11,137 | 0.909 |
| Agricultural and veterinary sciences | 10,347 | 3,186 | 0.650 |
| Civil engineering* | 4,790 | 3,773 | 0.254 |
| Industrial and information engineering | 32,109 | 4,865 | 1.320 |
| Total | 191,673 | 38,539 | 0.995 |

*Table 1: Average research output per capita indexed in WoS (articles, article reviews, and conference proceedings) in Italian universities between 2004 to 2008.*

*\* This discipline includes a number of fields of architecture for which the WoS does not serve as a significant census of research output.*

## 3. Methodology

To analyze the effects of size on research productivity we must first identify an indicator of output and then the means for its measurement. Since the main objective of research activity is to produce scientific advancement we require an indicator that can represent such advancement, and so naturally the observation of some form of codification. Essentially there are two techniques for measurement of the extent of scientific advancement: peer review and bibliometrics. The first depends on evaluation by experts in the areas of research outputs. However the costs and times of execution restrict its application to only a portion of the entire scientific output of organizations, and thus this approach is inadequate to respond to the research question in the present work. Bibliometrics identifies the citations of publications as proxy of impact on scientific advancement, notwithstanding the possible distortions implicit in this indicator (Moed, 2005; Glanzel, 2008). The main limit of citation analysis is that it is only applicable to those fields where publications indexed in WoS or Scopus provide a significant representation of the overall research output. Such significance is attained for most fields of the hard sciences (Moed et al., 2004, Abramo et al., 2009). A consequence of the limits is that for those fields not in the hard sciences it is impossible to test for the effects of size on research productivity (a limitation which would apply to both peer review and bibliometrics). However for the hard sciences it is possible to employ bibliometrics for the measurement of scientific productivity. The technique further permits measurement through the impact on scientific advancement, (citations), rather than just through the simple count of number of publications. The field of observation for our analysis consists of all Italian universities active in the hard sciences over the five years 2004 to 2008. Research productivity is measured by bibliometric techniques, essentially by an impact indicator linked to citations (see Section 3.2). The following subsections briefly describe the main characteristics of the Italian higher education system relevant to the analysis, followed by the methodological details.

### 3.1 The Italian higher education system

A description of the context of the study helps to understand the assumptions applied in our measurement model. In Italy, the Ministry of Education, Universities and Research (MIUR) recognizes a total of 95 universities as having the authority to issue legally-recognized degrees. With only rare exceptions these are public universities, largely financed by non-competitive government allocations. Until 2009, core government funding was input oriented, meaning distributed to universities in a manner intended to equally satisfy the needs of all, in function of their size and activities. The share of this core funding relative to total university income is now being reduced, declining from 61.5% in 2001 to 55.5% in 2007 (MIUR, 2009). It was only following the first national research evaluation exercise (VTR), conducted between 2004 and 2006, that a minimal share, equivalent to 7% of MIUR financing[4], was attributed in function of the assessment of research and teaching quality. Government planning is that this share will increase continually over the coming years.

New personnel can only enter the university system through public examinations, and career advancement also requires such examinations. Salaries are regulated at the nationally centralized level, calculated according to role (administrative, technical, or professorial), rank within role (for example: assistant, associate or full professor), and seniority. No part of the salary for professors is related to merit: wages are increased annually according to the government-established parameters. All professors are contractually obligated to carry out research, thus all universities are research universities: "teaching-only" universities do not exist. Each research staff member is classified as belonging to one specific disciplinary sector (SDS), of which there are 370[5], grouped in 14 University Disciplinary Areas (UDAs). The analysis in this work will be conducted at the level of these SDSs: the relation between research productivity and size is analyzed for each SDS in which every university is active. Since the teaching load for all scientists is established by law (350 hours per year) it can be considered as essentially uniform among all universities for the various SDSs, or at least undifferentiated by size class.

---

[4] Since MIUR financing composes 55.5% of the total, the share distributed on the basis of the VTR represents 3.9% of total income.

The specifics of the Italian higher education system rend it particularly suitable for the purposes of the present study, especially with reference to the typical challenges of modeling and of robustness in analysis, as mentioned in Section 1. Because the teaching load is similarly distributed across the Italian university sector, we can analyze the sole activity of research and carry out research productivity measurement alone. The total absence of competitive mechanisms tends to negate development of top universities, meaning that performance variability between universities is much less than that observed within them (Abramo et al., 2011). The large number of research organizations observed, together with the lack of concentration of top or unproductive scientists in a limited number of universities (as we show below), controls for the potential effect of different quality of scientists clustering in differently-sized universities. Further, the manner of government resource allocation does not permit any competitive advantage for individual universities in terms of production factors, including labor. For this reason, in the Italian system, analysis of returns to size could be considered a synonym of returns to scale, as larger labor size corresponds to a proportionate larger allocation of production factors other than labor. Furthermore, small, medium and large universities are distributed in a sufficiently even manner over the territory as to be able to exclude potential distortions in the size-productivity link due to localization effect. Finally the SDS classification of Italian research staff allows analyses that account for the possibility that production functions will vary depending of the field of research.

**3.2 Dataset and indicators**

As suggested, the unit of analysis is the SDS: measures of productivity are applied to the research staff of each university active in the national SDS. Data on staff members of each university and their SDS classification is extracted from the database on Italian university personnel, maintained by the Ministry for Universities and Research[6]. The bibliometric dataset used to measure productivity is extracted from the Italian Observatory of Public Research (ORP)[7], a database developed and maintained by the authors and derived under license from the Thomson Reuters WoS. Beginning from the raw data of the WoS, and applying a complex algorithm for reconciliation of the author's affiliation and disambiguation of the true identity of the authors, each publication (article, review and conference proceeding) is attributed to the university scientist or scientists that produced it (D'Angelo et al., 2010).

---

[5] The complete list is accessible on http://cercauniversita.cineca.it/php5/settori/index.php. Last accessed on July 20, 2011.

[6] http://cercauniversita.cineca.it/php5/docenti/cerca.php. Last accessed on July 20, 2011.

[7] www.orp.researchvalue.it. Last accessed on July 20, 2011.

To ensure the representativity of publications as proxy of the SDSs' research output, the field of observation was limited to those SDSs in the hard sciences[8] where at least 50% of Italian scientists produced at least one publication in the period 2004-2008. In the 183 SDSs thus examined, for this period there were 39,508 scientists (Table 2) distributed in 77 universities (out of a total of 60,000 research staff in all fields and all 87 universities).

| University disciplinary area | N. of SDSs | Universities | Research staff |
|---|---|---|---|
| Mathematics and computer sciences | 9 | 64 | 3,515 |
| Physics | 8 | 61 | 2,873 |
| Chemistry | 11 | 59 | 3,603 |
| Earth sciences | 12 | 48 | 1,439 |
| Biology | 19 | 66 | 5,785 |
| Medicine | 47 | 55 | 12,196 |
| Agricultural and veterinary sciences | 28 | 48 | 3,153 |
| Civil engineering | 7 | 49 | 1,455 |
| Industrial and information engineering | 42 | 68 | 5,489 |
| Total | 183 | 77 | 39,508 |

*Table 2: Universities and research staff with at least one publication, by UDA; data 2004-2008*

Measurement of research productivity of staff members of SDS $s$ of University $i$ is calculated by dividing the impact indicator "fractional standardized citations" (FSC), by the average number of the SDS research staff in the period 2004-2008.

The measure of FSC is based on the citations received for the publications by the scientists of the university SDS: citations of each publication are standardized by the median of citations[9] for all Italian publications of the same year and WoS subject category[10]. Italian universities are not homogenous in terms of fields of research investigation. Standardizing citations of each publication is thus not alone sufficient to avoid distortions in productivity assessment, given the varying intensity of publications in the various disciplines and fields[11]. To account for that, we carried out the analysis at SDS level[12]. Standardized citations are attributed to a university SDS in function of the co-authors on staff in that SDS relative to the total of co-authors. For the publications in the so-called "life science" categories (corresponding to 66 SDSs of 183), different weights are given to each co-author according to his/her position in the list and the character of the co-authorship (intra-mural or extra-mural)[13]. In formulae:

---

[8] In the Italian academic system, the hard sciences are matched in nine UDAs: mathematics and computer sciences; physics; chemistry; earth sciences; biology; medicine; agricultural and veterinary sciences; civil engineering; industrial and information engineering.
[9] Observed as of 30/06/2009.
[10] Standardizing citations to the median value rather than to the average, as frequently observed in literature, is justified by the fact that distribution of citations is highly skewed (Lundberg, 2007).
[11] The average number of publications per year by a physicist is 2.3 times what a mathematician produces.
[12] Although members of the same SDS may publish in different WoS subject categories their publication rate is not greatly affected by this. Differences in SS continue to reflect differences in productivity.
[13] Because in the life sciences, the different position in the authors' list of publications reflects the different contribution of the authors to the work, the following algorithm has been proposed by Italian scientists in the life sciences. It can be adapted to reflect different national contexts. If first and last authors belong to the same university, 40% of citations are attributed to each of them; the remaining 20% are divided among all other authors. If the first two and last two authors belong to different universities,

$$FSC_{i,s} = \sum_{j=1}^{N_{i,s}} \frac{C_j}{\bar{C}_j} \cdot n_{j,i,s}$$

With:

$C_j$ = number of citations received by publication $j$

$\bar{C}_j$ = median of citations received by all Italian publications of the same year and subject category of publication $j$

$n_{j,i,s}$ = fraction of authors of university $i$ and SDS $s$ on total coauthors of publication $j$, considering (if publication $j$ falls in life science subject categories) the position of each author and the character of the co-authorship (intra-mural or extra-mural).

$N_{i,s}$ = total number of publications authored by research staff in SDS $s$ of university $i$.

The productivity of SDS $s$ of university $i$ ($P_{i,s}$) is thus:

$$P_{i,s} = \frac{FSC_{i,s}}{RS_{i,s}}$$

With:

$RS_{i,s}$ = average research staff in university $i$ and SDS $s$, in the observed period

With this procedure, the measure of productivity is free of distortion due to any variations in publication-citation rates for the different fields of research.

## 4. Analysis and results

The analysis of co-authored publications in the hard sciences reveals that except for rare exceptions (research based on very large projects, such as in high energy and particle physics) the very large majority of such publications are co-authored by few authors. Meanwhile, progress in information and communications technologies, the advent of the Internet above all, reduction in transportation costs, and sharpening competition at the local and global levels, with resulting needs for specialization (Hara et al., 2003; Newell and Sproull,1982) are all clearly contributing to increasing adoption of international and/or extramural collaboration in research activity (Adams et al., 2005; Wagner and Leydesdorff, 2005; Zitt and Bassecoulard, 2004). Thus considering the share of extramural coauthors, the numbers of coauthors for each publication that belong to the same organization and field, except for rare exceptions, should result as being quite low. The expectation for the effects of personnel size on scientific productivity for an organizing is that, if it were to occur, it could be over very low size ranges: in the large majority of research fields we would thus expect increasing returns to size up to a few units of research and then constant beyond. Interference in the analyses could result from the presence of returns to scope, which would be probable in multidisciplinary types of research. The authors have recently begun a study of this phenomenon, however it is a particularly challenging task.

### 4.1 Associations of size with concentrations of top or unproductive scientists

Before verifying returns to size we must establish that top and unproductive scientists are not concentrated in particular sizes of universities. If more productive researchers were concentrated in large or small universities and unproductive ones in the opposite size, then a positive/negative correlation between productivity and size would not be sufficient proof of the occurrence of increasing/decreasing returns to size.

---

30% of citations are attributed to first and last authors; 15% of citations are attributed to second and last author but one; the remaining 10% are divided among all others. This algorithm has been proposed by Italian scientists in the life sciences. It can be adapted to reflect different national contexts.

The phenomenon is possible, though unlikely in such a broad field of observation.

The unproductives certainly exist. Observation is that 6,640 (16.8%) of the 39,512 research staff in the hard sciences did not publish any article in WoS-indexed journals over the period, and another 3,070 researchers (7.8% of total) produced at least one publication but did not receive citations. This means that 9,710 faculty (24.6% of total) did not have any impact on scientific progress (FSC nil). Moreover, 23% of university research staff produced 77% of overall scientific progress (an almost perfect fit to the Pareto principle). Therefore, in this section, the tests conducted hypothesize that the average productivity of a research group would be particularly sensitive to the presence of these unproductive researchers or to presence of top scientists, meaning to scientists with nil or very high FSC. With "top" we refer more precisely to scientists who, within their SDS, place above the 80[th] percentile for national rank of FSC, while "inactive" scientists are defined as having nil FSC.

The analysis is conducted by likelihood ratio Chi-square test (Williams, 1976), applied to two-way frequency tables, and Kendall's coefficient tau-b ($\tau_b$) of concordance. As a first step, the variables are dichotomized as seen in Table 3. For "size" and "inactive scientists", the values for subdivision are the relevant national medians of each SDS, while for low/high presence of "top scientists" the determining value is the figure of 20%[14].

| Variable | Dichotomization | |
|---|---|---|
| | Class | Values |
| Size: research staff of the university in the SDS | Small | ≤ Median of SDS |
| | Large | > Median of SDS |
| Top scientists: percentage of top scientists on total research staff in the SDS | Low | ≤ 20% |
| | High | > 20% |
| Non active scientists: percentage of inactive scientists on total research staff in the SDS | Low | ≤ Median of SDS |
| | High | > Median of SDS |

*Table 3: Dichotomization of variables for analysis of association of size of research unit with scientific quality of its staff members*

We provide the example of the Mechanics and machine design SDS, in the Industrial and information engineering UDA. There are 35 active universities in the SDS: Table 4 shows the distributions for university size-concentration of top scientists.

| | | Size | | |
|---|---|---|---|---|
| | | Small | Large | Total |
| Top scientists | Low | 11 *(11)* | 11 *(11)* | 22 |
| | High | 6 *(6)* | 7 *(7)* | 13 |
| | Total | 17 | 18 | 35 |

*Table 4: Two-way frequency table of size and top scientist concentration in Mechanics and machine design (expected values in brackets)*
*likelihood-ratio chi2: 0.0484*
*Kendall's $\tau_b$: 0.0372*
*p-value: 0.8260*

---

[14] With individual top scientists defined as those positioning above 80[th] percentile of national performance, a value of "top scientists" greater than 20% at a university indicates that their percentage in the staff is higher than the national average.

In this case there is no significant association (p-value = 0.8260) between concentration of top scientists and size. Frequencies in the two-way table are equal to those expected (in parentheses) under hypothesis of independence. This is further confirmed by the Kendall's $\tau_b$ coefficient of concordance, calculated at +0.0372[15].

The results for all the 183 SDSs are presented in Table 5, grouped by UDA. Only 32 SDSs (17% of overall total) show a significant association, with p-value 10%. The maximum percentage of significant cases is seen in the Civil engineering UDA (28% of SDSs), while in absolute numbers, the maximum is in Industrial and information engineering, with 10 significant SDSs (24% of total). In the Physics UDA there is no SDS where share of top scientists seems associated with university size.

| University disciplinary area | N° significant SDSs |
|---|---|
| Mathematics and computer sciences | 1 out of 9 (11%) |
| Physics | 0 out of 8 (0%) |
| Chemistry | 2 out of 11 (18%) |
| Earth Sciences | 3 out of 12 (25%) |
| Biology | 2 out of 19 (10%) |
| Medicine | 8 out of 47 (17%) |
| Agriculture and Veterinary Sciences | 4 out of 28 (14%) |
| Civil engineering | 2 out of 7 (28%) |
| Industrial and information engineering | 10 out of 42 (24%) |
| Total | 32 out of 183 (17 %) |

*Table 5: Summary by UDA of association between faculty size and top scientists concentration*

As previously indicated, the performance of research groups could also be influenced by greater/lesser incidence of low producing or inactive researchers. A second analysis thus detects any correlation between concentration of inactive scientists and size. The results are shown in Table 6. Of the 183 SDSs, only 33 (18%) show a significant link between the variables, with p-value 10%. Once again, the Civil engineering UDA has the greatest number of significant SDSs (43%) and the Earth sciences UDA shows no significant cases.

Intersection of the two above analyses indicates that in 59 of the total 183 SDSs, any correlation between average productivity and size of a research group might be attributed not to variable returns to size, but to quality of the labor production factor. Vice versa, in the 124 SDSs where distribution of quality of researchers is substantially uniform by size class, any differences in productivity can be attributed to occurrence of returns to size. In the next section we test for such differences.

| University disciplinary area | N° significant SDSs |
|---|---|
| Mathematics and computer sciences | 1 out of 9 (11%) |
| Physics | 2 out of 8 (25%) |
| Chemistry | 1 out of 11 (9%) |
| Earth Sciences | 0 out of 12 (0%) |
| Biology | 4 out of 19 (21%) |
| Medicine | 9 out of 47 (19%) |
| Agriculture and Veterinary Sciences | 3 out of 28 (11%) |
| Civil engineering | 3 out of 7 (43%) |
| Industrial and information engineering | 10 out of 42 (24%) |

---

[15] Kendall's tau-b assumes values between -1 (perfect inversion) and +1 (perfect concordance), with a value of 0 in cases of absence of association.

Total                               33 out of 183 (18%)

*Table 6: Summary by UDA of measures of association between staff size and concentration of inactive scientists*

### 4.2 Estimating returns to size

Examinations for returns to size as seen in the literature have been done by parametric or non-parametric approach. One of the more common non-parametric approaches is Data Envelopment Analysis (DEA)[16], which assigns a score for production efficiency in function of the distance from the efficiency frontier, under hypotheses of constant or variable returns to size. Any difference between scores under the two hypotheses, with a sufficiently high number of observations, is evidence of occurrence of variable returns (Avrikan, 2001). However the DEA approach does not allow for the real possibility that the university's production process could exhibit increasing returns at low input level and constant returns at intermediate and high levels. It is also very sensitive to outliers and requires a high number of observations. LOESS regression is also possible, although it cannot be applied under low number of observations, as seen in detailed analysis by field where datasets are likely limited in size (Bonaccorsi and Daraio 2005). Parametric methodologies require *a priori* assumptions concerning the form of the production function, preferably to be specified field by field. They also typically require a high number of observations, but there are never more than 60 universities per Italian SDS: this explains why preliminary tests by the authors showed a poor fit between many of the most common production functions and the Italian dataset.

We thus opt for non-parametric methodologies. To avoid potential distortions in some fields due to the low number of observations and to outliers we select dependence analysis methodology to quantify the relationship between productivity and size, through dichotomization. Specifically, to show occurrence of returns to size through dependence analysis, we again apply the Chi-square association test and tau-b index. We present the analysis for two example SDSs: one with no significant association between size and productivity and one with a link at significant level. As in the previous section we give the two-way frequency tables with association and concordance statistics. For dichotomization of size we use the same criteria as the previous section, while the threshold for productivity is the national median for the SDS: universities with productivity higher than median are classified as "high" productivity and universities under the median are "low" productivity.

We also provide box plots of absolute values of productivity in function of size, with size subdivided by quartiles. This permits a check on robustness of the dichotomization.

Table 7 shows the example of the Numerical analysis SDS, of the Mathematics and computer science UDA. The SDS has 47 active universities, substantially equi-distributed for the two classes of productivity and size. The likelihood ratio is not significant (p-value 0.308), indicating the absence of association between productivity and size.

|  |  | Size | | Total |
|---|---|---|---|---|
|  |  | Small | Large |  |
| Productivity | Low | 14 *(12)* | 10 *(12)* | 24 |
|  | High | 10 *(12)* | 13 *(11)* | 23 |

---

[16] DEA methodology seems particularly suited to comparing efficiency of research institutions, especially with the increasing availability of quantitative indicators for input and output. Abramo et al. (2011) is an example of a field-standardized application to national assessment.

|           | Total | 24 | 23 | 47 |

*Table 7: Two-way frequency table of size and productivity in the Numerical analysis SDS*
*likelihood-ratio chi2: 1.0409*
*p-value: 0.308*
*Kendall's τ$_b$: 0.1486*

The box plots in Figure 1 show that within each size quartile the distribution of productivity varies remarkably, with occurrence of outliers in the first three quartiles. The median values, represented by the central line for each box, remain substantially invariable with size. Further, the four size groups are similar for level of average productivity (NPC test p-value: 0.548), thus confirming that size has no correlation with productivity of the universities in this SDS.

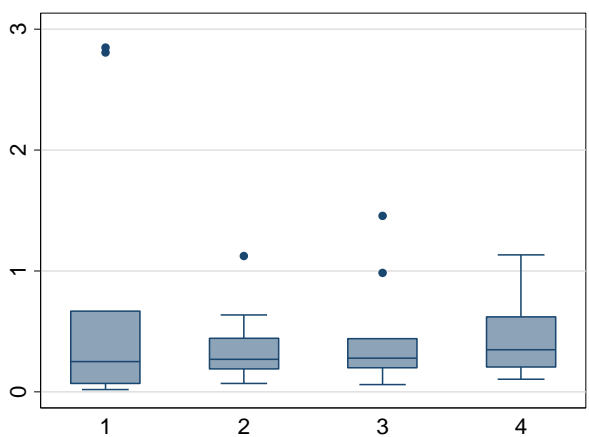

*Figure 1: Box plots for distributions of productivity by size quartile in the Numerical analysis SDS*
*NPC test p-value: 0.548*

The second example is the Organic chemistry SDS of Chemistry UDA. The 48 universities active in this SDS show concentrations in the small-size/low-productivity and large-size/high-productivity classes (Table 8). In this case, the Chi-square value is 3.0321, with significance 10% (p-value = 0.082) and the Kendall's coefficient tau-b value is +0.2500, indicating positive concordance, although not particularly high.

|  |  | Size | | |
|---|---|---|---|---|
|  |  | Small | Large | Total |
| Productivity | Low | 15 *(12)* | 9 *(12)* | 24 |
|  | High | 9 *(12)* | 15 *(12)* | 24 |
|  | Total | 24 | 24 | 48 |

*Table 8: Two way frequency table of size and productivity in the Organic chemistry SDS*
*likelihood-ratio chi2: 3.0321*
*p-value: 0.082*
*Kendall's τ$_b$: 0.2500*

The slight concordance is confirmed by the box plots in Figure 2. There is a step up in medians for absolute values of productivity between the first and second size quartiles. The average values of productivity also vary by size in a significant manner (p-value NPC test = 0.028), thus confirming occurrence of increasing returns to size in the Organic chemistry SDS.

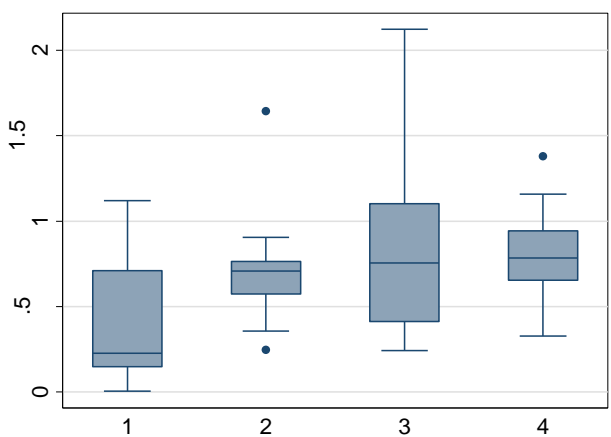

*Figure 2: Box plots of distributions of productivity by size quartile in the Organic chemistry SDS*
*p-value NPC test: 0.028*

This methodology was repeated for all 183 SDSs, initially disregarding the analysis of association seen in Section 4.1. Table 9 numbers the SDSs with significant association (p-value < 0.1) between size and productivity.

| University disciplinary area | N° significant SDS |
|---|---|
| Mathematics and computer sciences | 2 out of 9 (22%) |
| Physics | 0 out of 8 (0%) |
| Chemistry | 2 out of 11 (18%) |
| Earth Science | 2 out of 12 (17%) |
| Biology | 6 out of 19 (31%) |
| Medicine | 8 out of 47 (17%) |
| Agriculture and Veterinary Sciences | 7 out of 28 (25%) |
| Civil engineering | 5 out of 7 (71%) |
| Industrial and information engineering | 9 out of 42 (21%) |
| Total | 41 out of 183 (22%) |

*Table 9: Summary by UDA for measures of association between size and productivity*

Of all SDSs, 22% show potential increasing returns to size. The UDA with greatest number of significant SDSs (71%) is Civil engineering, while in Physics, none of the eight SDSs show a significant association between productivity and size.

Next we exclude those SDSs where the analysis of Section 4.1 shows that association between productivity and size could be due to concentration of top or unproductive researchers by size class of university. Following this step there remain only 18 of the 41 SDSs listed in Table 9, which we now present in Table 10. For these SDSs, the values of tau-b, all positive, demonstrate the occurrence of increasing returns to size. The Kendall's coefficients of concordance vary from a minimum of +0.2500 for Organic chemistry to a maximum of +0.7321 for Forestry and silviculture, both significant at 5%. Among the 18 fields that show increasing returns to size, the average Kendall's tau-b value is 0.3978, with a difference between minimum and maximum values of 0.4821 (variability not particularly high). It is notable that among these 18 SDSs, the four with highest tau-b are all in the Agriculture and veterinary science UDA.

| University disciplinary area | SDS | $\tau_b$ |
|---|---|---|

| Mathematics and computer sciences | Probability and mathematical statistics* | +0.2982 |
| --- | --- | --- |
| Chemistry | Analytical chemistry** | +0.4416 |
| | Organic chemistry* | +0.2500 |
| Earth Sciences | Paleontology and pale ecology* | +0.3333 |
| | Anthropology* | +0.3333 |
| Biology | Molecular biology* | +0.3050 |
| | Histology** | +0.4281 |
| Medicine | Odonto-stomalogical diseases* | +0.2941 |
| | Eye diseases* | +0.2782 |
| | Applied medical sciences * | +0.3352 |
| Agriculture and Veterinary sciences | Agricultural mechanics** | +0.5778 |
| | Inspection of food products of animal origin* | +0.4663 |
| | Infectious diseases of domestic animals** | +0.7321 |
| | Clinical veterinary medicine** | +0.7143 |
| Civil engineering | Maritime hydraulic construction and hydrology** | +0.3536 |
| | Science and technology of materials* | +0.2816 |
| Industrial and information engineering | Engineering and management* | +0.3095 |
| | Electronic and information bioengineering** | +0.4273 |

*Table 10: SDSs with significant increasing returns to size, grouped by UDA*
*\*\* p-value 0.05*
*\* p-value 0.10*

In summary, among the 124 SDSs in which correlation between productivity and size would not be affected by co-linearity of the two variables with "quality" of labor factor, there are only 18 cases (less than 15% of total) that show increasing returns to size. To make our findings more robust we carried out LOESS regression in the above 18 SDSs. Findings were aligned to the previous ones in most instances. In those SDSs where alignment did not occur, we noticed that it was due to outliers. Repeating LOESS regression without those outliers the results shown alignment again. This is presented in Figure 3 and in Figure 4 for the Electronic and information bioengineering SDS. The Kendall's coefficients of concordance of 0.4273 supports the idea of increasing returns to size, while the LOESS regression function of Figure 3 is flat, suggesting constant returns. Only after taking out the three outliers circled in Figure 3, we obtain a regression function showing increasing returns to size (Figure 4). Furthermore, we carried out LOESS regression on a sample of SDSs with constant returns to scale, according to dependence analysis. Taking out outliers, once again results were aligned with those from dependence analysis.

We can thus conclude that in the Italian university system, in 106 SDSs there are no strong or general links between size of university and level of production. There are 18 fields where such relation is significant and they are scattered among many disciplines. In 59 SDSs it is not possible to investigate the relationship.

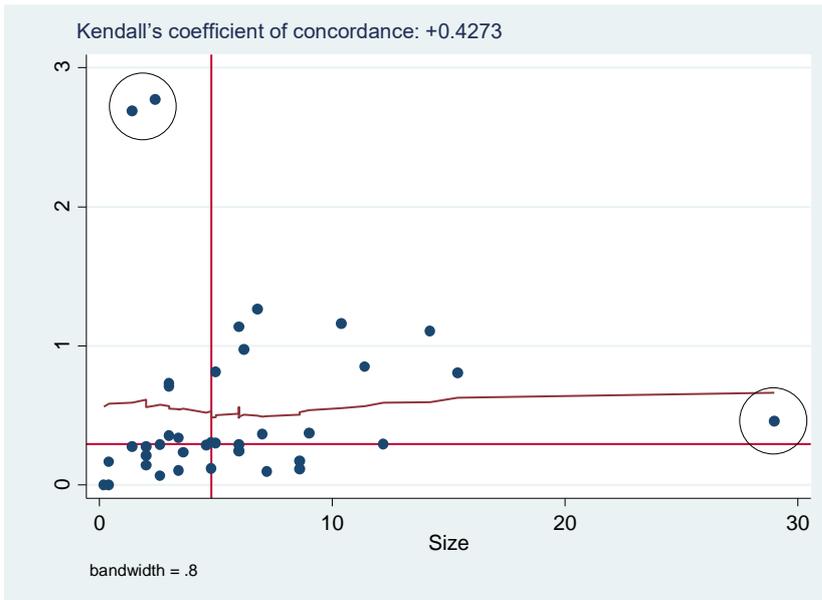

*Figure 3: LOESS regression of productivity vs. size in the Electronic and information bioengineering*

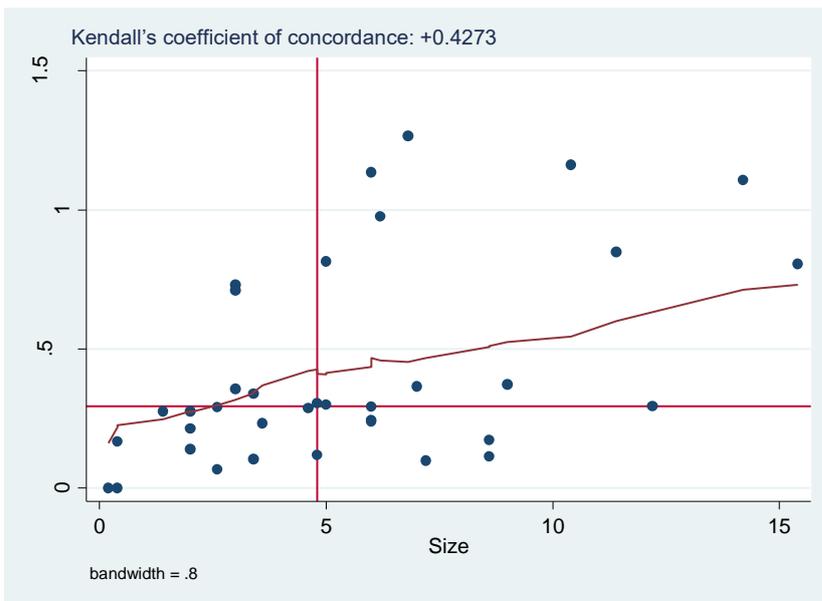

*Figure 4: LOESS regression of productivity vs. size in the Electronic and information bioengineering without outliers*

## 5. Conclusions

This study has examined the correlation between productivity of a research group and its size, in order to determine whether or not there are variable returns to size in research activity. The answer is relevant both to overall research system policy and management of institutions, but in spite of the questions' importance the literature does not provide mature and satisfactory response. The few studies which have attempted to address the issues suffer from a variety of critical problems. The method applied here attempts to overcome the problems by: i) isolating the "size" effect from that of other variables that can determine productivity (particularly the quality of the research staff); ii) providing a robust and innovative measure of productivity with analysis at the level of single fields of research; iii) analyzing an entire university system at field level (183 fields), with characteristics suitable to verification of the hypothesis (nation of Italy).

An analysis of concentrations of top/inactive researchers in the various universities demonstrates homogenous distribution of quality in research staff among institutions of various sizes, in over two thirds (124) of the fields investigated.

The subsequent analyses demonstrate that 106 fields of research are largely characterized by constant returns to size. The few significant cases of fields with increasing returns to size (18) are uniformly distributed, without concentration in the broader disciplines.

The specifics of the Italian higher education system made our investigation probably easier than it would be in other nations. To what extent our findings can be extended to other countries is something we cannot predict, although we cannot think of any plausible reasons why they should be different. The impression of the authors is that the production function in research is not country dependent, rather field dependent. Findings are aligned to expectations. Although collaboration rates in research are growing, the increasing abatement of communication barriers (especially with the advent of internet), makes cross-country and cross-organization collaborations easier and easier. In the case research projects require a large number of research staff, this can belong to different organizations and countries, making the size of the organization less critical.

This work was stimulated by two considerations: that the existence of returns to size could offer specific justification for adjustment to staff size in planning research organization systems, and would also require sorting of the organizations by size in the conduct of national research assessment exercises. The results of the analysis negate any such considerations in most fields.